\documentclass[%
 reprint,
superscriptaddress,
 citeautoscript,
 amsmath,amssymb,
 aps,
 prb,
]{revtex4-1}

\usepackage{graphicx}
\usepackage{dcolumn}
\usepackage{bm}
\usepackage{xcolor}
\usepackage{hyperref}
\hypersetup{
    colorlinks=true,
    linkcolor=blue,
    citecolor=blue,
    filecolor=blue,      
    urlcolor=blue,
    }

\newcommand{\Neqref}[1]{Eq.~\eqref{#1}}
\newcommand{\FS}[1]{\left\langle #1 \right\rangle_{\scriptscriptstyle\mathrm{FS}}}
 
\newcommand{\kb}{k_\mathrm{B}}
\newcommand{\tc}{T_\mathrm{c}}
\newcommand{\tco}{T_\mathrm{c0}}

\newcommand{\RR}{\mathbf{R}}
\newcommand{\vF}{\mathbf{v}_\mathrm{F}}
\newcommand{\pF}{\mathbf{p}_\mathrm{F}}
\newcommand{\ps}{\mathbf{p}_\mathrm{s}}

\newcommand{\NF}{\mathcal{N}_\mathrm{F}}

\begin{document}

\title{Thermopower and thermophase in a $d$-wave superconductor}

\author{Kevin Marc Seja}
 \affiliation{Department of Microtechnology and Nanoscience - MC2,
Chalmers University of Technology,
SE-41296 G\"oteborg, Sweden}
\author{Louhane Jacob}
 \affiliation{Department of Microtechnology and Nanoscience - MC2,
Chalmers University of Technology,
SE-41296 G\"oteborg, Sweden}
\affiliation{ENS de Lyon, 69342 Lyon, France}
\author{Tomas L\"ofwander}
 \affiliation{Department of Microtechnology and Nanoscience - MC2,
Chalmers University of Technology,
SE-41296 G\"oteborg, Sweden}

\date{\today}

\begin{abstract}
In an unconventional superconductor the interplay of scattering off impurities and Andreev processes may lead to different scattering times for electron-like and hole-like quasiparticles. Such electron-hole asymmetry appears when the impurity scattering phase shift is intermediate between the Born and unitary limits and leads to an expectation for large thermoelectric effects.
Here, we examine the thermoelectric response of a $d$-wave superconductor connected to normal-metal reservoirs under a temperature bias using a fully self-consistent quasiclassical theory. 
The thermoelectrically induced quasiparticle current is cancelled by superflow in an open circuit set-up, but at the cost of a charge imbalance induced at the contacts and extending across the structure. We investigate the resulting thermopower and thermophase and their dependencies on scattering phase shift, mean free path, and interface transparency. For crystal-axis orientations such that surface-bound zero-energy Andreev states are formed, the thermoelectric effect is reduced as a result of locally reduced electron-hole asymmetry.
For a semiballistic superconductor with good contacts we find thermopowers of order several ${\mu} V/K$, suggesting a thermovoltage measurement as a promising path to investigate thermoelectricity in unconventional superconductors.   
\end{abstract}

\maketitle

\section{Introduction}

The interest in thermal currents and thermoelectric effects in superconductors has been revived in recent years,
partly because the response to temperature gradients in superconductors is phase coherent \cite{maki_entropy_1965,guttman_phase-dependent_1997,guttman_interference_1998,Eom1998,zhao_phase_2003,Bezuglyi2003,zhao_heat_2004,Giazotto2012}.
In addition, heat management in nanoscale devices and circuits operating at low temperature is important in a wide range of applications in, e.g., thermometry, refridgeration, radiation detection \cite{giazotto_opportunities_2006}, and quantum technologies \cite{fornieri_towards_2017}.

The presence of a temperature gradient in a bulk superconductor leads to quasiparticle heat flow, reduced by the presence of the superconducting gap. In addition, an intricate thermoelectric response appears, where the thermoelectrically induced normal current is cancelled by counter superflow \cite{gin44}. The usual Seebeck effect was therefore initially expected to vanish. Efforts were instead made to pick up the magnetic flux created in a ring geometry by the counter superflow \cite{gal74,har74,zav74,har80}. But the experiment is complicated, for instance by the temperature dependence of the penetration depth, and the results have to some extent been controversial \cite{Shelly2016}. On the other hand, it was shown \cite{Artemenko1976} that in a device geometry with normal metals coupled to the superconductor, a charge imbalance signal could be induced by the temperature gradient. 
In conventional superconductors, however, the thermoelectric effect and the contact thermopower are small, since they require electron-hole asymmetry. Such asymmetry is very small in most metals because the density of states near the Fermi level is approximately constant and, in addition, the scattering time is to a good approximation energy independent. Later, it was suggested \cite{Pethick1979}, and experimentally confirmed \cite{Clarke1979}, that the application of a supercurrent in addition to the temperature gradient leads to an enhanced electron-hole asymmetry and a large thermopower \cite{Schoen1981}.

More recently, there are theories and experiments on engineering a large electron-hole asymmetry in superconducting heterostructures.
Examples range from injecting supercurrent into normal metals in multi-terminal geometries\cite{virtanen_thermoelectric_2007,chandrasekhar_thermal_2009}, over utilizing the spin degree of freedom in ferromagnet-superconductor hybrid devices \cite{machon_nonlocal_2013,ozaeta_predicted_2014,kalenkov_electron-hole_2014,giazotto_very_2015,kolenda_observation_2016,Hwang2018,Savander2020}, to injecting supercurrent into topological materials with edge states\cite{Blasi2020, BlasiDec2020}. Thus, the thermoelectric effect in conventional superconductors and heterostructures has been an active area of research. In contrast, much less is known about thermoelectric effects in unconventional superconductors.

In $d$-wave superconductors the electron-hole symmetry can be naturally broken when impurities scatter electron-like and hole-like quasiparticles differently \cite{monien87a,arf88,arf89,sal96,Lofwander2004Jul}. Such asymmetry is induced if the disorder is of certain type where the impurities scatter with a scattering phase shift $\delta_0$, in between the Born ($\delta_0$ small) and unitary ($\delta_0\rightarrow\pi/2$) limits.
The interplay of scattering off the impurity and Andreev processes leads to the formation of an impurity resonance state \cite{balatsky_impurity-induced_2006}. The resonance is at an energy within the maximum of the $d$-wave gap and is determined by the scattering phase shift. Within a homogeneous scattering model, where the effect of a dilute concentration of impurities is taken into account, an impurity band is formed. In thermal conductivity measurements the so-called universal limit has been found \cite{lee93,graf96,tai97}, showing the importance of the impurity band in transport. In addition to the impurity band in the density of states, electron-hole asymmetric scattering rates appear, which leads to the possibility of large thermoelectric effects that to the best of our knowledge have not been measured experimentally. Note that the electron-hole asymmetry we are considering here is only present in the superconducting state, and we will ignore any asymmetry in the band-structure or normal state scattering times.\cite{Gourgout2021}
From a fundamental point of view this type of thermoelectric response can be used to probe the influence of impurities in unconventional superconductors. But it has further implications for applications of unconventional superconducting devices, as for instance based on the high-$T_\mathrm{c}$ cuprates \cite{Trabaldo2019}.

In this work we consider the thermoelectric effect in $d$-wave superconductors with impurities scattering with an intermediate phase shift \cite{Lofwander2004Jul,lofwander_low-temperature_2005}. Building on our recent, fully self-consistent calculations of charge flow and charge imbalance in superconducting devices with normal metal leads \cite{Seja2021Sep}, we study a $d$-wave superconductor between two normal metal reservoirs with a temperature bias $\Delta T$, see Fig.~\ref{fig:model_tempbiased_wire}. We compute the thermopower, i.e., the thermoelectrically induced voltage $\Delta V$ due to the temperature bias $\Delta T$, in an open-circuit geometry where the charge current is zero. The induced voltage leads to a charge imbalance extending into the bulk of the superconductor and is large by the combined effects of the electron-hole asymmetric impurity scattering and the abundance of quasiparticle states around the nodes of the superconducting order parameter. The associated quasiparticle current flow is canceled everywhere by the moving condensate. As a result there is also a net superconducting phase difference, a thermophase, induced across the superconductor.

The paper is organized as follows. In Section~\ref{sec:model} we give the main assumptions of the model, while in Section~\ref{sec:methods} we outline the quasiclassical theory. A few details of the numerical procedures are given in Section~\ref{sec:scheme}. The results are collected in Section~\ref{sec:results}, and the final Section~\ref{sec:summary} gives a summary and discussion.

\section{Model and Methods}
\subsection{Model}\label{sec:model}
\begin{figure}[h!]
    \centering
    \includegraphics[width=0.9\columnwidth]{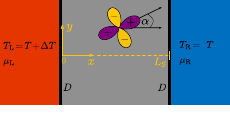}
    \caption{Principle setup of our model: A $d$-wave superconductor (grey) is connected to two reservoirs via contacts with a transmittance $D$ (black lines). The left, ''hot`` reservoir is at temperature $T_\mathrm{L} = T + \Delta T$ and chemical potential $\mu_\mathrm{L}$, while the right ''cold`` reservoir is at $T_\mathrm{R} = T$ and chemical potential $\mu_\mathrm{R}$. The angle $\alpha$ specifies the crystal-axes misalignment.}
    \label{fig:model_tempbiased_wire}
\end{figure}

The model we use to study thermoelectric effects in $d$-wave superconductors is depicted in Fig.~\ref{fig:model_tempbiased_wire}. A superconducting film is connected to two normal-metal reservoirs, one on the left side and one on the right side. Here, we assume that both reservoirs are connected via insulating barriers of equal transparency $D$. The left reservoir is kept at a temperature $T_L = T + \Delta T$, while the the right is at $T_R = T$. Experimentally, the temperature $T$ corresponds to the temperature of the cryostat that the hybrid structure is placed in, and one reservoir --- in our case the left --- is heated.

The temperature bias across the structure results in the injection of quasiparticles from the reservoirs into the superconductor and transport in the superconducting $ab$-plane ($xy$-plane). The film is considered homogeneous in the perpendicular $c$-axis direction.
In addition, the temperature difference is assumed to be applied from left to right in a homogeneous fashion in the transverse $y$-direction. The thermoelectric response will then also be translationally invariant in the transverse direction and we may compute the flow of heat and charge along the $x$-direction only. Since the charge current is zero everywhere there is no induced magnetic field and the vector potential is zero.

We consider system sizes $L_\mathrm{S}$ smaller than the inelastic scattering length $\ell_\mathrm{in}$, while the elastic mean free path $\ell$ can take arbitrary values. As we discussed in some detail in Ref.~\onlinecite{Seja2021Sep}, this means that the dwell time for injected non-equilibrium quasiparticles is short compared to time scales of inelastic scattering processes in the superconductor,\cite{Eliashberg1971,Kaplan1976,Catelani2010} and final relaxation takes place far inside the normal metal contact reservoirs. Here we consider relatively small nano-devices with $L_\mathrm{S}<100\xi_0$, where this can be met.

We assume that a static temperature difference between left and right sides can be upheld. In addition, we assume that any transient response present in a real experiment during the establishment of this temperature difference has been damped out by dissipation in the reservoirs. In this case we may consider the long-time limit non-equilibrium stationary state.

\subsection{Methods}\label{sec:methods}

The steady-state nonequilibrium response of our structure to the external temperature bias is obtained by solving the time-independent quasiclassical equation of motion 
\begin{equation}
i \hbar \vF \cdot \nabla \check{g} + \left[ \varepsilon \hat{\tau}_3 \check{1} - \check{h}, \check{g} \right] = 0,
\label{eq:transportequation}
\end{equation}
where $[\check A,\check B]$ denotes a commutator between matrices $\check A$ and $\check B$. In addition to Eq.\eqref{eq:transportequation}, the quasiclassical Green's function $\check{g}$ has to satisfy the normalization condition
$\check{g}^2 = - \pi^2 \check{1}$. These equations were originally derived by Eilenberger\cite{Eilenberger1968Apr} and independently by Larkin and Ovchinnikov,\cite{Larkin1969} and later generalized to the nonequilibrium case by Eliashberg \cite{Eliashberg1971}.
The $\check{~}$ denotes that $\check{g}$ is a matrix in Keldysh space,
\begin{equation}
\check{g}(\pF,\RR,\varepsilon) =
\begin{pmatrix}
\hat{g}^\mathrm{R}(\pF,\RR,\varepsilon) & \hat{g}^\mathrm{K}(\pF,\RR,\varepsilon)
\\
0 & \hat{g}^\mathrm{A}(\pF,\RR,\varepsilon)
\end{pmatrix}.
\label{eq:gcheckDefinition}
\end{equation}
In the following, we will omit the explicit reference to the dependencies on momentum direction $\pF$, coordinate $\RR$, and energy $\varepsilon$ if it does not lead to confusion. The spectrum is determined by the retarded (R) and advanced (A) components while the distribution function is related to the Keldysh (K) propagator.
All three elements in Eq.~\eqref{eq:gcheckDefinition} carry a $\hat{~}$ to indicate that they are matrices in particle-hole (Nambu) space. We denote Pauli matrices in Nambu space $\hat\tau_i$ ($i=1,2,3$), and the third, $\hat\tau_3$, enters in \Neqref{eq:transportequation}. The retarded and advanced components have the form
\begin{align}
\hat{g}^{\mathrm{R},\mathrm{A}} 
=
\begin{pmatrix}
g & f\\
\tilde f & \tilde g
\end{pmatrix}^\mathrm{R,A},
% \mp 2\pi i \begin{pmatrix}
% \mathcal{G} & \mathcal{F}
% \\
% -\tilde{ \mathcal{F} } & - \tilde{ \mathcal{G} }
% \end{pmatrix}^{\mathrm{R},\mathrm{A}}
% \pm i \pi \hat{\tau}_3,
\label{eq:NambuSpaceGFsDef}
\end{align}
%where $\hat{\tau}_3$ is the Pauli-z matrix in Nambu space.
while the Keldysh component is
\begin{equation}
\hat{g}^\mathrm{K} =
\begin{pmatrix}
g^K & f^K\\
-\tilde f^K & -\tilde g^K
\end{pmatrix}.
% - 2 \pi i 
% \begin{pmatrix}
% \mathcal{X} & \mathcal{Y}
% \\
% \tilde{\mathcal{Y}} & \tilde{\mathcal{X}}
% \end{pmatrix}^\mathrm{K}.
\label{eq:NambuSpaceGKsDef}
\end{equation}
The diagonal components are propagators, while the off-diagonal anomalous propagators encode superconductivity. Particle-hole conjugation gives a relation
\begin{align}
\tilde{A}(\pF,\RR,\varepsilon) = A^*(-\pF,\RR,-\varepsilon^*),
\label{eq:TildeSymmetry}
\end{align}
between tilde and ''non-tilde`` objects.
The elements of $\hat{g}^\mathrm{R,A,K}$ are formally matrices in spin space, but we assume a spin-degenerate system and spin singlet $d$-wave superconductivity.

The self-energies taken into account are the mean-field superconducting order parameter as well as scalar impurities,
\begin{equation}
\check{h}(\pF,\RR,\varepsilon) = \check{h}_\mathrm{mf}(\pF,\RR) + \check{h}_\mathrm{s}(\RR,\varepsilon).
\end{equation}
The singlet $d$-wave order parameter is
\begin{equation}
\Delta(\pF,\RR)=\Delta_0(\RR)\eta_d(\pF)i\sigma_2,
\end{equation}
where $i\sigma_2$ is the singlet spin structure ($\sigma_2$ is the 2nd Pauli matrix in spin space). The $d$-wave orbital basis function is $\eta(\pF)=\sqrt{2}\cos\left[2\left(\varphi_\mathrm{F}-\alpha\right)\right]$, where $\varphi_\mathrm{F}$ is the angle between $\pF$ and the $x$-axis and the angle $\alpha$ gives the misorientation of the $d$-wave clover to the device main axis, see Fig.~\ref{fig:model_tempbiased_wire}. 
The order parameter amplitude satisfies the following gap equation
\begin{equation}
\Delta_0(\RR) = \lambda\NF\int_{-\varepsilon_\mathrm{c}}^{\varepsilon_c}\frac{d\varepsilon}{8\pi i}
\FS{\mbox{Tr}\left[i\sigma_2\eta_d(\pF) f^\mathrm{K}(\pF,\RR,\varepsilon)\right]},
\label{eq:gapequation}
\end{equation}
where the trace over spin projects out the singlet component, $\lambda$ is the superconducting coupling constant, $\varepsilon_\mathrm{c}$ is an energy cut-off, and $\NF$ is the density of states per spin in the normal state. The linearized gap equation can be used to eliminate the coupling constant $\lambda$ and the cut-off $\varepsilon_\mathrm{c}$ in favor of the superconducting transition temperature $T_\mathrm{c}$, which then becomes the natural energy scale of the theory.
We assume a circular Fermi surface, in which case the Fermi surface average appearing in \Neqref{eq:gapequation} is defined as
\begin{equation}
\FS{\dots} = \int_0^{2\pi}\frac{d\varphi_\mathrm{F}}{2\pi}(\dots).
\end{equation}
The Keldysh matrix structure of $\check{h}_\mathrm{mf}$ is simple,
$\check{h}_\mathrm{mf} = \hat\Delta\check{1}$,
while the Nambu structure is $\hat\Delta=\Re(\Delta)\hat\tau_1-\Im(\Delta)\hat\tau_2$. In equilibrium, the order parameter can be taken real, while under non-equilibrium it is complex. We may introduce the phase $\chi(\RR)$ of the order parameter as $\Delta_0(\RR)=|\Delta_0(\RR)|\exp[{i\chi(\RR)}]$. Its gradient gives the superfluid momentum $\ps(\RR)=\tfrac{\hbar}{2}\nabla\chi(\RR)$ and is related to the presence of superflow.

Assuming an average dilute impurity concentration $n_\mathrm{i}$, the impurity self-energy is found from the t-matrix equation in the non-crossing approximation \cite{AGD:book}:
\begin{align}
\check{h}_s = n_\mathrm{i} \check{t} \equiv n_\mathrm{i}
\begin{pmatrix}
\hat{t}^R & \hat{t}^K
\\
0 & \hat{t}^A
\end{pmatrix}.
\label{eq:selfEnergyTMatrixEquation}
\end{align}
For scattering that is isotropic in momentum space with an $s$-wave scattering potential $u_0$ the elements of $\check{t}$ satisfy the equations
\begin{align}
\hat{t}^\mathrm{~R,A} &= \frac{u_0 \hat{1} + u_0^2 \NF \FS{\hat{g}^\mathrm{R,A}}}{\hat{1} - \left[  u_0 \NF \FS{\hat{g}^\mathrm{R,A} }\right]^2   },
\\
\hat{t}^\mathrm{K} &= \NF \hat{t}^\mathrm{R} \FS{\hat{g}^\mathrm{K}} \hat{t}^\mathrm{A}.
\label{eq:tKDefinition}
\end{align}
We will express the two free parameters of this scattering model, $n_\mathrm{i}$ and $u_0$, in terms of the scattering energy $\Gamma_u$ and scattering phase shift $\delta_0$, 
\begin{align}
\Gamma_u &\equiv \frac{n_\mathrm{i}}{\pi \NF},
\\
\delta_0 &\equiv \arctan (\pi u_0 \NF) .
\end{align}
The so-called pair-breaking energy is then given by
\begin{align}
\Gamma \equiv \Gamma_u \sin^2 \delta_0,
\label{eq:PairBreakingParameter}
\end{align}
which is also related to the normal-state mean free path
\begin{align}
\ell = \frac{\hbar v_\mathrm{F}}{2\Gamma}.
\label{eq:mfpGammaRelation}
\end{align} 
The natural length scale of superconducting phenomena is the superconducting coherence length, defined as
\begin{align}
\xi_0 \equiv \frac{\hbar v_\mathrm{F}}{2 \pi \kb T_\mathrm{c 0}}.
\label{eq:xi0Definition}
\end{align}
The properties of an unconventional superconductor typically depend on the ratio between mean free path and the superconducting coherence length $\ell/\xi_0$.
For $d$-wave superconductors, scalar impurities are pair breaking and the mean free path is bounded by a critical mean free path, $\ell_\mathrm{c} \approx 3.6\xi_0$, below which the superconducting order parameter vanishes \cite{xu95}.

In terms of these parameters, the retarded impurity self-energy takes the form
\begin{align}
    \hat{h}_\mathrm{s}^\mathrm{R}(\RR,\varepsilon) &= \Gamma_u
    \frac{\sin\delta_0\cos\delta_0\hat{1}+\sin^2\delta_0\tfrac{1}{\pi}\FS{\hat{g}^\mathrm{R}(\pF,\RR,\varepsilon)}}{\cos^2\delta_0+\sin^2\delta_0\left(\tfrac{1}{\pi}\FS{\hat{g}^\mathrm{R}(\pF,\RR,\varepsilon)}\right)^2}\nonumber\\
&\equiv
\begin{pmatrix}
\Sigma_\mathrm{s}^\mathrm{R}(\RR,\varepsilon) & \Delta_\mathrm{s}^\mathrm{R}(\RR,\varepsilon)
\\
\tilde\Delta_\mathrm{s}^\mathrm{R}(\RR,\varepsilon) & \tilde\Sigma_\mathrm{s}^\mathrm{R}(\RR,\varepsilon)
\end{pmatrix},
\label{eq:ImpRetarded}
\end{align}
with an analogous expression for the advanced self-energy $\hat{h}_\mathrm{s}^A$.
By tuning $\delta_0$ we vary the character of scattering between the two limiting cases of the weak-scattering Born limit, $\delta_0 \rightarrow 0,~\Gamma_u \rightarrow \infty$, with $\Gamma$  constant, and the strong-scattering unitarity limit, $|\delta_0| \rightarrow \pi/2$. The first term proportional to the unit-matrix in Nambu space quantifies the amount of electron-hole asymmetry induced by the impurities \cite{monien87a,arf88,arf89,sal96,Lofwander2004Jul}. In the Born and unitary limits, it vanishes, while for intermediate phase shifts it is finite. 
The Keldysh selfenergy $\hat{h}_\mathrm{s}^\mathrm{K}$ is obtained by combining Eqs.~\eqref{eq:selfEnergyTMatrixEquation} and \eqref{eq:tKDefinition}, and has the elements
\begin{align}
\hat{h}_\mathrm{s}^\mathrm{K}(\RR,\varepsilon) \equiv
\begin{pmatrix}
\Sigma^\mathrm{K}_\mathrm{s}(\RR,\varepsilon) & \Delta^\mathrm{K}_\mathrm{s}(\RR,\varepsilon)
\\
- \tilde{\Delta}^\mathrm{K}_\mathrm{s}(\RR,\varepsilon) & - \tilde{\Sigma}^\mathrm{K}_\mathrm{s}(\RR,\varepsilon)
\end{pmatrix}
.
\end{align}

\begin{figure}[t]
     \centering
     \includegraphics{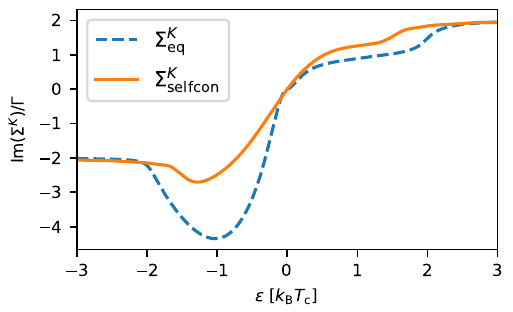}
     \caption{Imaginary part of the Keldysh self-energy $\Sigma^K$ in equilibrium (dashed blue) and nonequilbrium self-consistent (solid orange) for $\Delta T = 0.1 \kb \tc$, $T = 0.25 \kb \tc$, $D=1$, $\Gamma = 0.1\pi$, $\delta_0 = \pi/4$, computed at the center of the superconductor at $x=L_\mathrm{s}/2$. }
     \label{fig:SigmaK_selfcon}
 \end{figure}

Within the quasiclassical approximation, the normal state electron-hole asymmetry due to bandstructure is neglected. In contrast,
the electron-hole asymmetry in \Neqref{eq:ImpRetarded} is finite for $T<T_\mathrm{c}$ already in equilibrium, meaning that there is a bulk first-order (linear) thermoelectric response \cite{Lofwander2004Jul}. Here we go beyond linear response and compute the stationary non-linear thermopower in a device geometry. In Fig.~\ref{fig:SigmaK_selfcon} we show the energy-dependence of the diagonal component in Nambu space of the Keldysh self-energy both in equilibrium and non-equilibrium. The electron-hole asymmetric scattering rate is clear from $\Im\Sigma^\mathrm{K}(-\varepsilon)\neq -\Im\Sigma^\mathrm{K}(\varepsilon)$.
%%%%%%%%%%%%%%%%%%%%%%%%%%%%%%%%%%%%%%%%%%%%%%%%%%%%%%%%%%%%%%%%%%%%%%%%%%%

There is a freedom to parameterize the Keldysh Green's function $\hat{g}^K(\pF,\RR,\varepsilon)$ in terms of different distribution functions. As we discussed in Ref~\onlinecite{Seja2021Sep}, one choice that is appealing when interpreting results is to use the distribution matrix $\hat{f}(\pF,\RR,\varepsilon)$, 
\begin{align}
\hat{g}^\mathrm{K} = \hat{g}^\mathrm{R} \hat{f}  - \hat{f} \hat{g}^\mathrm{A}.
\label{eq:gKeldyshInH}
\end{align}
The matrix $\hat{f}(\pF,\RR,\varepsilon)$ itself can be written as
\begin{align}
\hat{f}
=
 f_1 \hat{1} + f_3 \hat{\tau}_3
=
\begin{pmatrix}
h & 0 
\\
0 & -\tilde{h}
\end{pmatrix}
.
\label{eq:h-splitting}
\end{align}
The two components, $f_1(\pF,\RR,\varepsilon)$ and $f_3(\pF,\RR,\varepsilon)$, become after averaging over momentum the energy mode, $\FS{f_1(\pF,\RR,\varepsilon)}$, and charge mode, $\FS{f_3(\pF,\RR,\varepsilon)}$, used in literature on non-equlibrium phenomena in diffusive $s$-wave superconductors \cite{Schmid1975Jul}. The relevant observables can be written in terms of the two momentum-dependent distributions.

The quasiparticle chemical potential has the form
\begin{equation}
\phi(\RR) = -\frac{1}{2e} \int\limits_{-\infty}^{\infty} \mathrm{d}\varepsilon \FS{f_3(\pF,\RR,\varepsilon) \mathcal{N}(\pF,\RR,\varepsilon)},
\label{eq:phi_hmatrix}
\end{equation}
the charge current is
\begin{align}
\mathbf{j}(\RR)\!=\!-e\NF\!\int\limits_{-\infty}^{\infty}\!\mathrm{d}\varepsilon
\FS{ \vF f_1(\pF,\RR,\varepsilon) \mathcal{N}(\pF,\RR,\varepsilon) },
\label{eq:ChargeCurrentDefinition_hmatrix}
\end{align}
and the energy current reads
\begin{align}
\mathbf{j}^\mathrm{th}(\RR) 
\!=\!-\NF\!\int\limits_{-\infty}^{\infty}\! \mathrm{d}\varepsilon~\varepsilon \FS{ \vF f_3(\pF, \RR, \varepsilon) \mathcal{N}(\pF, \RR,\varepsilon)}.
\label{eq:HeatCurrentDefinition_hmatrix}
\end{align}
We use $j_0=e v_\mathrm{F}\NF\kb\tc$ as the unit for charge current, and $j_0^\mathrm{th}=v_\mathrm{F}\NF (\kb\tc)^2$ as unit for energy current. 
In all of the above equations, $\mathcal{N}(\pF,\RR,\varepsilon)$ is a normalized, momentum-resolved local density of states per spin:
\begin{equation}
\mathcal{N}(\pF, \RR, \varepsilon) = -\frac{1}{4\pi}\mathrm{Im}~\mathrm{Tr}\left[\hat{\tau}_3 \hat{g}^R(\pF, \RR, \varepsilon)\right].
\end{equation}
It is related to the total local density of states via
\begin{equation}
N(\RR,\varepsilon) = 2\NF\FS{\mathcal{N}(\pF, \RR, \varepsilon)}.
\end{equation}
An alternative separation of the Keldysh Green's function is obtained by splitting the distribution function into a local-equilibrium distribution function
\begin{equation}
h^\mathrm{le}(\RR,\varepsilon) = \tanh\frac{\varepsilon-e\phi(\RR)}{2T},
\label{eq:h_le}
\end{equation}
with the base temperature $T$, and an anomalous part $h^\mathrm{a}$, such that
\begin{align}
h = h^\mathrm{le} + ( h - h^\mathrm{le}) \equiv h^\mathrm{le} + h^a.
\end{align}
With this choice, we have $\hat{g}^\mathrm{K}=\hat{g}^\mathrm{le}+\hat{g}^\mathrm{a}$, where the so-called anomalous part $\hat{g}^\mathrm{a}$ contains all the information of the non-equilibrium form of the distribution. The advantage, as we shall see below, is that quasiparticle flow due to the temperature gradient is mainly given by the anomalous part while the condensate response is mainly given by the local equilibrium term:
\begin{equation}
\mathbf{j}(\RR) = \mathbf{j}^\mathrm{le}(\RR) + \mathbf{j}^\mathrm{a}(\RR).
\label{eq:currentSplitting}
\end{equation}
Indeed, noting that the local equilibrium distribution in \Neqref{eq:h_le} is momentum-independent, when inserted in \Neqref{eq:ChargeCurrentDefinition_hmatrix}, the contribution to the Fermi-surface average comes from an imbalance in the local density of states for left-moving and right-moving states. For the negative-energy continuum, forming the condensate, the Doppler shifts $\vF\cdot\ps$ due to superflow self-consistently form the supercurrent that shows up in $\mathbf{j}^\mathrm{le}(\RR)$. 

In a non-equilibrium situation, a local temperature is not well defined since the energy mode is modified and there is a spatially dependent electrochemical potential. But following Ref.~\onlinecite{Heikkila}, we may define a spatially dependent effective temperature.
First we note that the energy mode in equilibrium is $f_1^\mathrm{eq}(\varepsilon,T)=\tanh(\varepsilon/2\kb T)$. From a Sommerfeld expansion one can show that
\begin{align}
\int\limits_0^\infty \mathrm{d}\varepsilon~\varepsilon \left[ f_1^\mathrm{eq}(\varepsilon, T=0) - f_1^\mathrm{eq}(\varepsilon, T) \right] = \frac{\pi^2}{6}\kb^2 T^2.
\end{align}
In non-equilibrium we then define the local effective temperature $T_\mathrm{eff}(\RR)$ as
\begin{align}
\kb^2 T_\mathrm{eff}(\RR)^2 \!\equiv \frac{6}{\pi^2}\int\limits_0^\infty \mathrm{d}\varepsilon\,\varepsilon\,
[ &~f_1^\mathrm{le}(\RR,\varepsilon,T=0)\nonumber\\
&- \FS{f_1(\pF,\RR,\varepsilon)}],
\label{eq:TeffAverage}
\end{align}
where $f_1^\mathrm{le}(\RR,\varepsilon,T=0)=1-\Theta[\varepsilon-e\phi(\RR)]-\Theta[\varepsilon+e\phi(\RR)]$ is the zero-temperature local equilibrium energy mode for local potential $\phi(\RR)$. 
Furthermore, we define left-mover and right-mover averages of $f_1$ and $f_3$ in Eq.~\eqref{eq:h-splitting} via
\begin{align}
\langle f_1 \rangle_{\substack{\rightarrow \\ \leftarrow} } = \frac{ \langle h \rangle_\pm - \langle \tilde{h} \rangle_\mp }{2},\quad
\langle f_3 \rangle_{\substack{\rightarrow \\ \leftarrow} } = \frac{ \langle h \rangle_\pm + \langle \tilde{h} \rangle_\mp }{2}.
\label{eq:modes_partialFSAverage}
\end{align}
Here we use a partial Fermi surface average
\begin{align}
\langle A \rangle_\pm &\equiv 
\int\limits_{0}^{2\pi}\!
\frac{d\varphi_\mathrm{F}}{\pi}
\!~ A(\varphi_\mathrm{F}) \Theta(\pm \cos \varphi_\mathrm{F}),
\label{eq:PositiveFermiAverage}
\end{align}
where the Heaviside step function $\Theta(\pm\cos\varphi_\mathrm{F})$ is unity if the projection of $\vec{v}_\mathrm{F}$ on the transport axis $\hat{x}$ is positive (+) or negative (-), and zero otherwise. The half-space averaged $f_1$ ($f_3$) in \Neqref{eq:modes_partialFSAverage} is an odd (even) function of energy, just as its full Fermi surface average and the corresponding longitudinal (transversal) mode in Usadel theory.
Note also that the normalization in \Neqref{eq:modes_partialFSAverage} is such that $\FS{f_i} = ( \langle f_i \rangle_\rightarrow + \langle f_i \rangle_\leftarrow)/2$.
We then obtain an effective temperature $T_{\mathrm{eff},\rightarrow}$ for right-movers by replacing the full Fermi-surface average in Eq.~\eqref{eq:TeffAverage} by $\langle f_1 \rangle_{{\rightarrow}}$ as defined in Eq.~\eqref{eq:modes_partialFSAverage}, with an analogous defintion for $T_{\mathrm{eff}, \leftarrow}$.

\subsection{Calculation scheme}\label{sec:scheme}
To solve Eq.~\eqref{eq:transportequation}, we use a parametrization of Eqs.~\eqref{eq:NambuSpaceGFsDef}-\eqref{eq:NambuSpaceGKsDef} in terms of coherence amplitudes\cite{Nagato1993, Schopohl1995, Schopohl1998} $\gamma$ and $\tilde{\gamma}$, and distribution functions $x$ and $\tilde{x}$\cite{eschrig_distribution_2000,Eschrig2009Oct}. Using $\mathcal{G}^R \equiv ( 1 - \gamma^\mathrm{R} \tilde{\gamma}^\mathrm{R})^{-1}$ and $\mathcal{F}^\mathrm{R} \equiv \mathcal{G}^\mathrm{R}\gamma^\mathrm{R}$, we have
\begin{align}
\hat{g}^\mathrm{R} = 
-2\pi i
\!
\begin{pmatrix}
\mathcal{G}& \mathcal{F}
 \\
-\tilde{\mathcal{F}} & -\tilde{\mathcal{G}}
\end{pmatrix}^\mathrm{R}
+ i \pi \hat{\tau}_3,
\end{align}
with an analogous definition for the advanced functions, while the Keldysh component takes the form
\begin{align}
\hat{g}^\mathrm{K}
\equiv
-2 \pi i 
\begin{pmatrix}
\mathcal{G} & \mathcal{F}
\\
-\tilde{\mathcal{F}} & -\tilde{\mathcal{G}}
\end{pmatrix}^\mathrm{R}
\begin{pmatrix}
x & 0
\\ 0 & \tilde{x}
\end{pmatrix}
\begin{pmatrix}
\mathcal{G} & \mathcal{F}
\\
-\tilde{\mathcal{F}} & -\tilde{\mathcal{G}}
\end{pmatrix}^\mathrm{A}.
\label{eq:riccati_x}
\end{align}
The parametrizing functions themselves satisfy a set of coupled transport equations. The coherence function $\gamma$ satisfies a Riccati equation,
\begin{align}
\left( i\hbar\vF\cdot\nabla + 2 \varepsilon \right) \gamma^{\mathrm{R},\mathrm{A}}
&= \bigl( \gamma \tilde{\Delta} \gamma + \Sigma \gamma - \gamma \tilde{\Sigma} - \Delta \bigr)^{\mathrm{R},\mathrm{A}}\!,
\label{eq:GammaEquation}
\end{align}
while the equation for the distribution function $x$ reads
\begin{align}
&i \hbar \vF \cdot \nabla x - \left[ \gamma \tilde{\Delta} + \Sigma \right]^\mathrm{R} x - x \left[ \Delta \tilde{\gamma} - \Sigma \right]^\mathrm{A}
\nonumber
\\
&=
- \gamma^\mathrm{R} \tilde{\Sigma}^\mathrm{K} \tilde{\gamma}^\mathrm{A} + \Delta^\mathrm{K} \tilde{\gamma}^\mathrm{A} + \gamma^\mathrm{R} \tilde{\Delta}^\mathrm{K} - \Sigma^\mathrm{K}.
\label{eq:xEquation}
\end{align}
These equations for $\tilde{\gamma}$ and $\tilde{x}$ can be obtained by the tilde symmetry, Eq.~\eqref{eq:TildeSymmetry}. The transport equations are solved from a start point to an end point along a trajectory direction specified by $\vF$. Analytic solutions to Eqs.~\eqref{eq:GammaEquation}-\eqref{eq:xEquation} can be found in a region of constant selfenergies\cite{Eschrig2009Oct}. We thus model our self-energy in the superconductor as piecewise constant in space and numerically calculate the analytic solution to Eqs.~\eqref{eq:GammaEquation}-\eqref{eq:xEquation} in order to propagate the relevant functions along the trajectory \cite{Grein2013}.
The functions in the normal reservoirs and the central superconducting region are connected via boundary conditions written in terms of scattering matrices for the normal metal-insulator-superconductor interfaces \cite{eschrig_distribution_2000,Eschrig2009Oct, Zhao2004}. In this paper, the scattering matrix, or equivalently the transparency $D$ in Fig.~\ref{fig:model_tempbiased_wire}, is assumed to be independent of the trajectory angle $\varphi_\mathrm{F}$.
The functions in the normal reservoirs entering the boundary conditions are assumed to have their normal metal reservoir forms. The coherence amplitudes are zero, $\gamma=\tilde\gamma=0$, while the distributions have equilibrium form, $x_\mathrm{L/R}=\tanh\left[(\varepsilon-\mu_\mathrm{L/R})/2\kb T_\mathrm{L/R}\right]$, in terms of the reservoir temperatures including the temperature bias as well as the thermoelectrically induced chemical potential difference, see below.
The parametrization of $\hat{g}^\mathrm{K}$ in terms of distribution functions $x$ and $\tilde x$ in \Neqref{eq:riccati_x} leads to an equation of motion, Eq.~\eqref{eq:xEquation}, that is easier to solve than the corresponding equations for the distributions $h$ and $\tilde h$.\cite{Eschrig2009Oct}
At the end of the calculations we may transform back from $x$($\tilde{x}$) to $h$($\tilde{h}$) as used in Eqs.~(\ref{eq:gKeldyshInH})-(\ref{eq:h-splitting}). The latter two functions facilitate interpretation of the stationary nonequilibrium in terms of the modes $f_1$ and $f_3$. For further details, we refer to Ref.~\onlinecite{Seja2021Sep}. 

Determining the thermopower of the superconductor requires finding the potential difference between the two reservoirs that develops as a result of the applied temperature bias.
The temperature bias leads to the injection of a nonequilibrium distribution into the superconductor. Starting from an equilibrium guess for all selfenergies we solve Eq.~\eqref{eq:transportequation} for $\hat{g}^\mathrm{R,A,K}$ and use the solutions to update all selfenergies. We can iteratively find the chemical potential of the two reservoirs via a fixed-point iteration of the form
\begin{align}
\mu_\mathrm{L(R)}^{(n+1)} = \mu_\mathrm{L(R)}^{(n)} - p~\frac{\kb\tco\, j_\mathrm{L(R)}}{j_0}, 
\label{eq:ChemicalPotentialIteration}
\end{align}
where $j_\mathrm{L(R)}$ is the current flow between the superconductor and the left (right) reservoir and $p$ is a numerical parameter of order unity. The iteration then converges towards $j_\mathrm{L(R)} = 0$. Physically, no current flows between the superconductor and the reservoirs in the steady state in an open-circuit setup. Inside the superconductor we require conservation of charge current, meaning in this case that the current everywhere is zero. This is guaranteed by solving for the self-energies and the $\phi$-potential self-consistently. The results shown here have $|j(\RR)| < 10^{-5}j_0$ everywhere. 

Once a selfconsistent solution is obtained, we obtain the thermopower 
\begin{align}
S = - \frac{\mu_\mathrm{L} - \mu_\mathrm{R}}{\Delta T}, %\frac{e}{\kb} 
\label{eq:thermopower}
\end{align}
where $\mu_\mathrm{L} - \mu_\mathrm{R}$ is the voltage drop across the structure and $\Delta T$ is the applied temperature bias.
\section{Results}\label{sec:results}
\begin{figure}[th]
    \centering
    \includegraphics{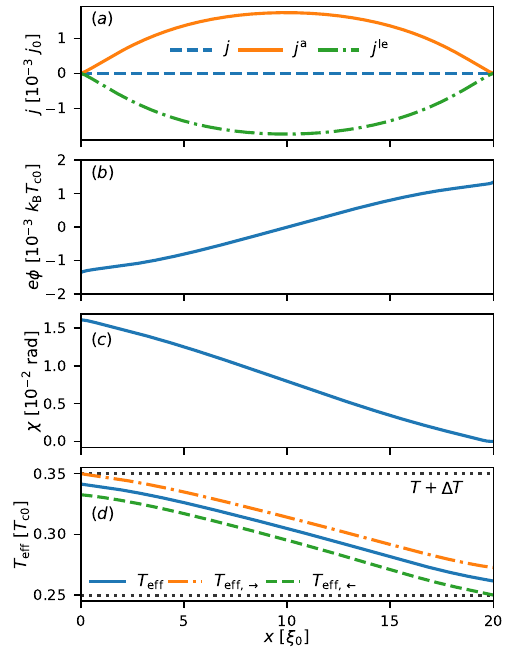}
    \caption{Main physical quantities for a temperature bias $\Delta T = 0.1 T_\mathrm{c0}$ and physical parameters given by $\delta_0 = 0.9$, $\Gamma=0.1\pi\kb\tco$, $L_\mathrm{S} = 20 \xi_0$, $D=1$, and $T = 0.25T_\mathrm{c0}$. (a) The anomalous, $j^\mathrm{a}$, and local-equilibrium, $j^\mathrm{le}$, currents across the superconductor. They add up to a total current, $j$, that is zero everywhere. (b) The quasiparticle chemical potential and (c) the phase drop throughout the structure. (d) The effective temperature $T_\mathrm{eff}$, as well as the right-mover and left-mover effective temperatures $T_{\mathrm{eff}, \rightarrow}$ and $T_{\mathrm{eff}, \leftarrow}$.}
    \label{fig:results_example}
\end{figure}

In Fig.~\ref{fig:results_example} we present the main physical quantities for a temperature bias of $\Delta T=0.1\tco$ and base temperature $T=0.25\tco$. The temperature bias leads to a thermal current from hot to cold $j^\mathrm{th}=1.45\cdot10^{-3}j_0^\mathrm{th}$, that in the absence of inelastic scattering is conserved across the device. The temperature gradient also leads to a thermoelectric response, a quasiparticle or so-called anomalous current, $j^\mathrm{a}(x)$, that also flows from hot to cold, see Fig.~\ref{fig:results_example}(a). In a bulk linear response calculation \cite{Lofwander2004Jul} we would have $j^\mathrm{a}=-\eta\partial_x T>0$, where $\eta$ is the thermoelectric response function ($\eta>0$ for $\delta_0>0$). 
Here we show the stationary non-linear response between left and right normal metal reservoirs. In the interior of the superconductor, the condensate moves, which shows up as a local equilibrium component, $j^\mathrm{le}(x)$. The total current is zero, $j(x)=j^\mathrm{a}(x)+j^\mathrm{le}(x)=0$, everywhere. In Fig.~\ref{fig:results_example}(b) we show the potential $\phi(x)$ that develops to ensure that the total current vanishes also at the interfaces to the reservoirs at $x=0$ and $x=L_\mathrm{S}$. For instance, at the left interface, the potential is negative, meaning that it suppresses $j^\mathrm{a}$ down to zero and no current is flowing out to the contact in the open circuit set-up. The counter superflow from right to left leads to that the phase $\chi(x)$ of the superconducting order parameter grows from the right lead to the left lead, see Fig.~\ref{fig:results_example}(c). As a result a phase difference,
\begin{equation}
\Delta\chi=\chi(0)-\chi(L_\mathrm{S}),
\end{equation}
also called thermophase, is formed in response to the temperature bias.

\begin{figure}[t]
    \centering
    \includegraphics{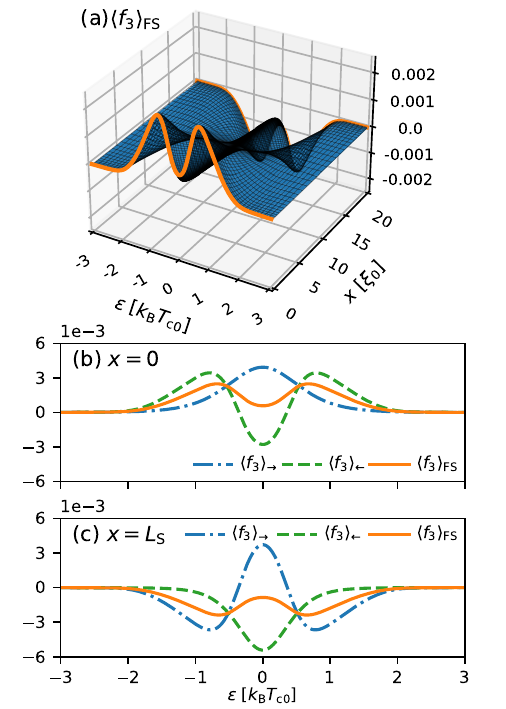}
    \caption{Example of the distribution $f_3$ for the same parameters as in Fig.~\ref{fig:results_example}. (a) Mode $\FS{f_3}$ across the structure. The orange curves are the shapes in the superconductor directly at the N-S interfaces. (b) Separation of $\FS{f_3}$ (solid orange) into components for right-movers ($\langle f_3 \rangle_\rightarrow$, dash-dotted blue) and left-movers $\langle f_3 \rangle_\leftarrow$ (dashed green) at the left N-S interface ($x=0$). (c) The same separation but at the right N-S interface ($x=L_\mathrm{S}$).}
    \label{fig:distribution}
\end{figure}
 Throughout the superconductor, the electron distribution has a non-equilibrium form. To quantify it we separate it into the right-moving and left-moving energy and charge modes. The energy mode $\FS{f_1(\pF,x,\varepsilon)}$ is close to a thermal distribution and the introduction of a local effective temperature $T_\mathrm{eff}(x)$ is illuminating, see Fig.~\ref{fig:results_example}(d). The effective temperatures of the right- and left-moving quasiparticles, $T_\mathrm{eff,\rightarrow}(x)$ and $T_\mathrm{eff,\leftarrow}(x)$, connect to the temperatures of the respective reservoirs of origin in the case of fully transmissive interfaces. For finite transparency, there is a jump between reservoir temperatures and right- and left-mover temperatures, determined by the thermal resistance of the barrier.
 The temperatures $T_\mathrm{eff,\substack{\rightarrow \\ \leftarrow}}$ are analogues of the right-moving and left-moving quasipotentials, $\phi_{\substack{\rightarrow \\ \leftarrow}}(x)$, appearing in a voltage bias set-up \cite{Seja2021Sep}. We note that the example in Fig.~\ref{fig:results_example} is for a rather dirty system. For cleaner systems, the difference between $T_\mathrm{eff,\rightarrow}(x)$ and $T_\mathrm{eff,\leftarrow}(x)$ is larger. 
 The charge mode $\FS{f_3(\pF,x,\varepsilon)}$ is presented Fig.~\ref{fig:distribution}. The non-monotonic energy dependence is due to the non-trivial energy-dependence of the scattering rate, given by the diagonal component $\Im\Sigma^K$ displayed in Fig.~\ref{fig:SigmaK_selfcon}.

\begin{figure}[t]
    \centering
    \includegraphics{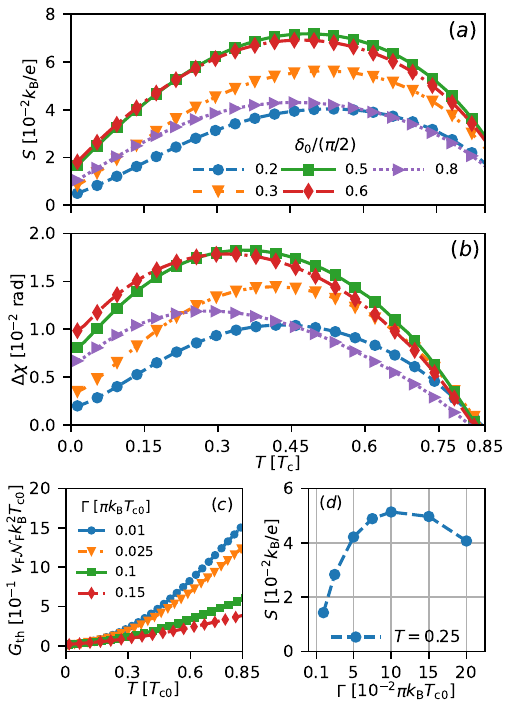}
    \caption{Base temperature dependence of
    (a) thermopower and (b) thermophase for a variety of scattering phase shifts for a fixed pairbreaking parameter $\Gamma=0.1\pi\kb\tco$ (mean free path $\ell=10\xi_0=0.5L_\mathrm{S}$). The markers in (b) correspond to the ones given in (a). 
    (c) Base temperature dependence of the thermal conductance for a variety of pairbreaking parameters. The phase shift is $\delta_0=0.9$ but the thermal conductance is largely phase shift independent.
    (d) Thermopower as function of pairbreaking parameter for fixed bath temperature $T=0.25\tco$ and $\delta_0=0.9$. The line is a guide to the eye.
    In all cases $D = 1$ and $\Delta T=0.1\tco$.
    }
    \label{fig:results_impurity_variation}
\end{figure} 

\begin{figure*}
    \centering
    \includegraphics{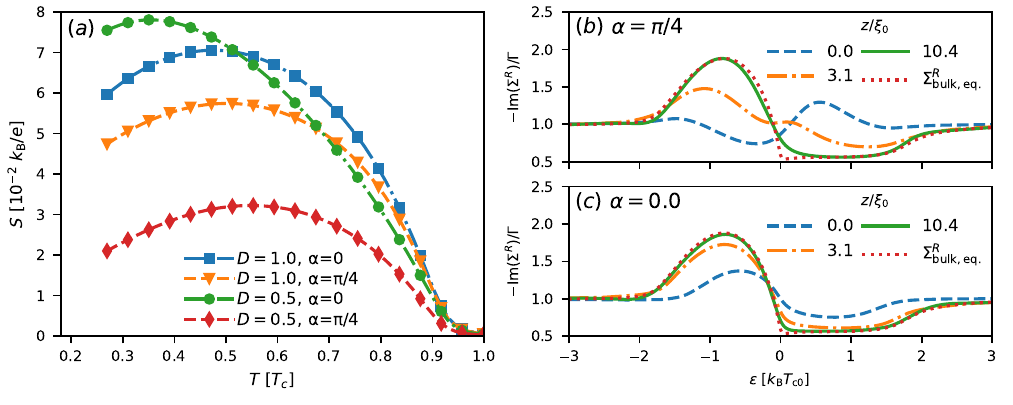}
    \caption{Influence of interface transparency $D$ and crystal-axes misalignment $\alpha$. (a) Thermopower $S$ for two different values of $D$ and two values of $\alpha$. For all cases we have $\delta_0 =0.9$, $\Gamma = 0.1\pi\kb\tco$, $L_\mathrm{S}=20\xi_0$, and $\Delta T = 0.1 T_\mathrm{c0}$.
    (b) Imaginary part of the equilibrium impurity self-energy $\Sigma^R$ at different spatial positions for $D=0.5$ and $\alpha=\pi/2$.
    (c) Imaginary part of the equilibrium impurity self-energy $\Sigma^R$ at different spatial positions for $D=0.5$ and $\alpha=0$. The comparison between (b) and (c) shows that Andreev bound states at the interface reduce the electron-hole asymmetry. As a result, the thermopower $S$ reduces in (a) for $D=0.5$ when $\alpha = \pi/4$.
     }
    \label{fig:results_thermopower_transp_and_alpha}
\end{figure*}

In Fig.~\ref{fig:results_impurity_variation} we present an overview of the base temperature dependence of observables for different impurity parameters. Only positive phase shifts are shown, since for negative $\delta_0$ both the thermopower and the thermophase change signs as the electron-hole asymmetry is inverted. Technically, this is due to the sign change of the first term in \Neqref{eq:ImpRetarded} proportional to the unit matrix in Nambu space. Since we have a finite temperature bias $\Delta T=0.1\tco$, the thermopower in Fig.~\ref{fig:results_impurity_variation}(a) ends at finite values. In the linear response case \cite{Lofwander2004Jul}, the thermoelectric coefficient $\eta$ vanishes linearly with $T$ as $T\rightarrow 0$, while here it remains finite since the hot reservoir injects quasiparticles at $T=\Delta T$. In the normal state, for $T>T_c$, the electron-hole asymmetry vanishes in our quasiclassical theory, and $S=0$. Therefore, as $T$ approaches $\tc$, the thermopower goes to zero. The thermopower as well as the thermophase is therefore the largest at intermediate temperatures. The phase-shift dependence reflects the amount of electron-hole asymmetry, and the thermopower and thermophase are the largest at phase shifts between the Born and unitary limits. On the other hand, the thermal conductance has a very weak dependence on the phase shift, and is instead limited by the mean free path $\ell$, see Fig.~\ref{fig:results_impurity_variation}(c). As a transport quantity, the thermal conductance is larger in ballistic systems.

In Fig.~\ref{fig:results_impurity_variation}(d) we show the dependence of the thermopower on the pairbreaking parameter $\Gamma$. The non-monotonic dependence can be understood as follows. 
For small values of $\Gamma$ an injected quasiparticle is unlikely to scatter while passing through the supercondutor since $\ell > L_\mathrm{S}$, and thus, the thermopower is small. For the intermediate range $\ell \approx L_\mathrm{S}$ the thermopower saturates and only weakly depends on $\Gamma$.
On the other hand, for large $\Gamma$, the energy-dependence of the self-energy seen in Fig.~\ref{fig:SigmaK_selfcon} is increasingly broadened and the electron-hole asymmetry is reduced, which reduces the thermopower. In addition, for diffusive devices, in the sense $\ell\ll L_\mathrm{S}$, the right-mover and left-mover distributions become more equal to each other which reduces the thermoelectric response. In summary, to maximize the thermopower the device should not be in the ballistic ($\ell\gg L_\mathrm{S}$) or diffusive ($\ell\ll L_\mathrm{S}$) limits, i.e., $L_\mathrm{S}$ should be comparable or larger than the mean free path $\ell$.

Lastly, we address how interface transparency and crystal-axes misalignment affect the thermovoltage. Fig.~\ref{fig:results_thermopower_transp_and_alpha}(a) shows $S(T)$ for two transparencies, $D=1$ and $D=0.5$, and two misaligment angles, $\alpha=0$ and $\alpha=\pi/4$. For $\alpha = 0$ and temperatures $T\gtrsim 0.5\tco$ reduced transparency leads to a reduction of the thermopower. Reduced transparency leads to back reflection of quasiparticles and a thermal resistance of the interface. The effective temperature then jumps across the interfaces and the temperature gradient is reduced in the interior superconductor and the thermoelectric effect is also reduced. On the other hand, for lower temperatures, $T\lesssim 0.5\tco$, the reduction of the transparency instead leads to an enhanced thermopower. This is a combination of two effects that become of importance at lower temperature. Firstly, in the case of $D=1$, the electron-hole asymmetry gets suppressed by the inverse proximity effect, where superconductivity is suppressed near the contacts, thereby reducing the thermopower. Secondly, in the case of $D=0.5$ the interface resistance is higher and a higher voltage is required to cancel the thermoelectrically induced current. These two effects compensate for the thermal resistance of the barrier and the thermopower is enhanced at $D=0.5$.

For $\alpha=\pi/4$ and $D=1$ we see a reduction in $S(T)$, as compared with $\alpha=0$. This is because we inject quasiparticles into the node of the $d$-wave order parameter where there are a lot of states. The device is more like a normal metal, with lower thermopower. For $\alpha=\pi/4$ and $D=0.5$, there is a large density of zero-energy Andreev bound states formed at the surface \cite{kashiwaya_tunnelling_2000,lofwander_andreev_2001}. Comparison of $\Sigma^R(x,\varepsilon)$ in Fig.~\ref{fig:results_thermopower_transp_and_alpha}(b) and (c) shows that the bound states effectively invert the bulk electron-hole asymmetry at the interface, and suppress the bulk asymmetry over a length scale of several coherence lengths. This leads to a further reduction of the thermopower.

\section{Summary and discussion}\label{sec:summary}
In this paper, we have studied the stationary nonlinear thermoelectric response of a $d$-wave superconductor connected to normal-metal reservoirs under a temperature bias. Earlier results \cite{Lofwander2004Jul} had predicted within linear response a large bulk thermoelectric effect caused by an impurity-induced electron-hole asymmetry for scattering phase shifts in between the Born and unitary limits. Using self-consistent quasiclassical theory for the stationary nonequilibrium response, we studied physical quantities such as the thermopower, the thermophase, and the nonequilibrium distribution in a device setup with normal metal leads. Aside from varying the base temperature of the setup, we examined the effects of varying interface transparency, crystal-axes misalignment, and mean free path.

Our results show that the thermoelectric response of the superconductor leads to a thermovoltage between the reservoirs, with the thermopower being maximal for $\delta_0 \sim \pi /4$. For such intermediate phase shifts, a system with good contacts and a mean free path on the order of the system size has a thermopower of $S \approx 5 \times 10^{-2}~\kb/e$. Taking $\mathrm{YBa}_2\mathrm{Cu}_3\mathrm{O}_{7-\delta}$ ($T_\mathrm{c} \approx 90$~K) as an example compound, an applied temperature difference of $\Delta T = 0.1 T_\mathrm{c} \approx 9 ~ \mathrm{K}$ results in a voltage drop of $\Delta V \approx -40 \mu V$, a value that is experimentally easily accessible.

Within our quasiclassical theory, any electron-hole asymmetry induced by bandstructure or normal state energy-dependent scattering times, are neglected. Instead the asymmetry we consider is induced in the superconducting state through impurity scattering, the interplay with Andreev processes, and the resulting impurity resonance states. There is a recent interest in the Seebeck effect in the non-superconducting state of high-temperature superconducting materials where superconductivity is suppressed by a high magnetic field \cite{Gourgout2021}. 
Since the electron-hole asymmetry is modified in the superconducting state, our results indicate that additional information about for instance impurity scattering in the material could be extracted in the superconducting state through a thermopower measurement between normal metal leads. The thermoelectric signal we find is sufficiently large that it could compete with other contributions already present in the normal state. For instance, different types of impurities, scattering with different signs of the scattering phase shift, would lead to opposite signs of the Seebeck effect in the superconducting state.

In conclusion, our results predict a measurable thermovoltage in normal-metal/$d$-wave superconductor hybrid structures. This suggests a voltage measurement in such structures as a promising complimentary experimental approach to the flux measurements on purely superconducting rings. Our conclusions should be valid in a broader perspective for other symmetries of the order parameter, as long as impurities lead to a large electron-hole asymmetry. Also in a conventional superconductor with an impurity band of Yu-Shiba-Russinov states at finite energies within the $s$-wave gap, a similarly large thermoelectric response coefficient has been predicted through a linear response calculation \cite{Kalenkov2012}, and our approach is valid in this case as well.
Lastly, we point out that our prediction of a large thermopower implies a non-negligible Peltier effect that could influence the charge-transport behavior of $d$-wave superconducting devices. We leave an examination of this effect to future studies. 

\begin{acknowledgments}
We thank J. Splettstoesser, T. Bauch, and A. Yurgens for valuable discussions.
We acknowledge financial support from the Swedish research council and financial support of the internship of L.J. from the Excellence Initiative Nano at Chalmers. The computations were enabled by resources provided by the Swedish National Infrastructure for Computing (SNIC) at NSC partially funded by the Swedish Research Council through grant agreement no. 2018-05973.
\end{acknowledgments}

\providecommand{\noopsort}[1]{}\providecommand{\singleletter}[1]{#1}%
\end{document}